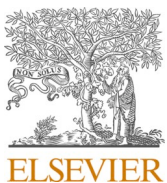
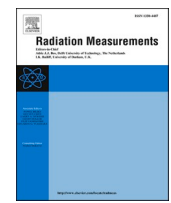

# Amorphous silicon detectors for proton beam monitoring in FLASH radiotherapy

Nicolas Wyrsch [a,*], Luca Antognini [a], Christophe Ballif [a], Saverio Braccini [b], Pierluigi Casolaro [b,c,d], Sylvain Dunand [a], Alexander Gottstein [b], Matt Large [e], Isidre Mateu [b], Jonathan Thomet [a]

[a] *Photovoltaics and Thin-Film Electronics Laboratory, Institute of Electrical and Microengineering (IEM), Ecole Polytechnique Fédérale de Lausanne (EPFL), Neuchâtel, Switzerland*
[b] *Albert Einstein Center for Fundamental Physics (AEC), Laboratory for High Energy Physics (LHEP), University of Bern, Bern, Switzerland*
[c] *Department of Physics "Ettore Pancini", Università degli Studi di Napoli Federico II, Napoli, Italy*
[d] *INFN Sezione di Napoli, Complesso Univ. Monte S. Angelo, Napoli, Italy*
[e] *Centre for Medical Radiation Physics, University of Wollongong, Wollongong, Australia*

ABSTRACT

Ultra-high dose rate radiation therapy (FLASH) based on proton irradiation is of major interest for cancer treatments but creates new challenges for dose monitoring. Amorphous hydrogenated silicon is known to be one of the most radiation-hard semiconductors. In this study, detectors based on this material are investigated at proton dose rates similar to or exceeding those required for FLASH therapy. Tested detectors comprise two different types of contacts, two different thicknesses deposited either on glass or on polyimide substrates. All detectors exhibit excellent linear behaviour as a function of dose rate up to a value of 20 kGy/s. Linearity is achieved independently of the depletion condition of the device and remarkably in passive (unbiased) conditions. The degradation of the performance as a function of the dose rate and its recovery are also discussed.

## 1. Introduction

Ultra-high dose rate radiation therapy (so-called FLASH radiotherapy) is attracting a lot of attention for cancer treatments (Vozenin et al., 2019; Wilson et al., 2020; Esplen et al., 2020). FLASH irradiations are characterised by the delivery of ultra-high dose rates (>40 Gy/s) in a fraction of a second. These dose rates are several orders of magnitude higher than conventional ones (~5 Gy/min) used clinically. This novel radiation delivery modality has been observed to reduce the toxicity to healthy tissues by maintaining the same anti-tumour efficacy as conventional radiotherapy. In external beam radiotherapy, including FLASH therapy, the use of proton beams offers several benefits (Hughes and Parsons, 2020; Matuszak et al., 2022). Thanks to the Bragg peak, proton beams allow for a targeted treatment of deep-seated tumours with reduced entrance dose and enhanced conformity. Currently, proton FLASH therapy is under investigation through pre-clinical and clinical experiments (Mascia et al., 2023; Diffenderfer et al., 2022). However, the typical irradiation conditions of FLASH therapy pose new challenges for dosimetry calling for new materials or detectors for dose monitoring (Romano et al., 2022; Jolly et al., 2020; Casolaro et al., 2022).

Hydrogenated amorphous silicon (a-Si:H) is known for being one of the most radiation-resistant semiconductors and, for this reason, it has been studied in the eighties and nineties as an active material for particle detection in high-energy physics experiments, but without much practical use (Wyrsch and Ballif, 2016). The amorphous nature of a-Si:H with a high content of hydrogen, which passivates intrinsic defects renders this material quite immune to radiation defects. With the need for more radiation-resistant detectors, a-Si:H, given its high radiation hardness, is attracting renewed attention, and recent studies demonstrated its performance for high-dose X-ray monitoring (Menichelli et al., 2023; Large et al., 2023; Grimani. et al., 2023). These results, thus, indicate a high potential for a-Si: H-based devices for dose monitoring in cancer radiation therapy, including FLASH and hadron therapy.

Typical a-Si:H detectors consist of a thick intrinsic layer playing the role of the active volume, sandwiched between thin doped n- and p-layers deposited on a variety of substrates coated with a metallic contact. Another contact (metal or conductive oxide) is deposited as a top electrode. This type of devices thus offers a well-defined interaction volume that can be extremely small (practically down to a few tens of cubic micrometers); charge collection takes place only in between the top and bottom contacts where an electric field is present. Alternatively, n- and p-layers can be replaced by electron- and hole-selective contacts (charge-selective contacts) using, for example, respectively, $TiO_2$ or $MoO_x$. Such materials offer the possibility to fabricate 3D detectors but also offer a potential for higher sensitivity for planar devices (Menichelli et al., 2021).

In this contribution, the sensitivity and radiation resistance of a-Si:H devices of two different thicknesses with doped and charge-selective contacts are characterised and compared. We especially focus on the linearity of such detectors at ultra-high dose rates. In view of applications to proton FLASH therapy, devices were characterised up to dose rates two orders of magnitude larger than those used in conventional radiotherapy settings to insure a linear behaviour in the range of interest. In addition, devices deposited on glass and polyimide (PI) were also compared. Deposition on PI offers the possibility to fabricate flexible detectors which could be of interest in real-time in-vivo dosimetry.

## 2. Experimental details

### 2.1. Samples

For the present investigation, several sets of a-Si:H detectors were fabricated either on Schott AF32 glass (500 μm thick) or on Al-coated PI (PIT1N-ALUM from capLINQ, 25.4 μm thick) substrates. All substrates were first coated by sputtering with an Al/Cr layer stack used as the detector back contact. The devices were then deposited either in the n-i-p configuration or with selective contacts in the $TiO_2$-i-$MoO_x$ configuration with 2 different thicknesses (2 and 10 μm) for the intrinsic i- layer. The latter was deposited for all samples at ca. 200 °C by plasma-enhanced chemical vapor deposition (PECVD). For selective contact configurations, a ca. 13 nm thick sputtered $TiO_2$ layer used as selective electron contact and a ca. 10 nm thick sputtered $MoO_x$ layer used as selective hole contact were deposited by sputtering. A top 65 nm thick Indium Tin Oxide (ITO) layer was deposited by sputtering through a shadow mask on all devices to define top contacts with an active area of 5x5 $mm^2$. To limit peripheral leakage in n-i-p devices, the i-p stack layer was partially removed by dry etching. More experimental details on the fabrication of the devices can be found in (Wyrsch et al., 2004). Finally, the samples were attached to a PCB, and the top contacts were connected through wires glued with a graphite paste to wire pins for electrical connections. Some of the samples, as well as a schematic cross-section of the devices, are pictured in Fig. 1. Note that due to severe delamination of some of the thick samples, only thick samples on glass could be characterised.

### 2.2. Proton beamline

For the proton irradiation, the research beam line of the cyclotron (IBA Cyclone 18/18) installed at the Bern University Hospital (Inselspital) (Bracini, 2013) was used. The main usages of this 18 MeV proton accelerator are the production of radionuclides, especially $^{18}F$ for Positron Emission Tomography (PET) imaging and multi-disciplinary research performed by means of a specific beam transfer line (BTL) ending in a second bunker. For the present study, the beam size on the sample was defined by a collimator (dump ring in Fig. 2 (Anders et al., 2022)). The beam was extracted in air through a 50 μm Al foil. The fluence on the sample placed in front of the Al foil was evaluated from the measured current on the collimator by considering the open area of the collimator (we used either 1 × 1 $cm^2$ or 2 × 2 $cm^2$ open area). The beam uniformity upstream of the collimator was evaluated with the UniBEaM, a two-dimensional beam monitor based on scintillating optical fibres developed by the Bern group (Potkins et al., 2017). The proton flux's calibration is verified both with an ionization chamber (PTW Freiburg GmbH, Monitor Chamber Type 786) and with radiochromic film dosimetry (Casolaro et al., 2019). The calibration is verified weekly and has been recently used for experiments on radiation hardness (Anders et al., 2022), testing of novel detectors (Braccini et al., 2023), and radiobiology.

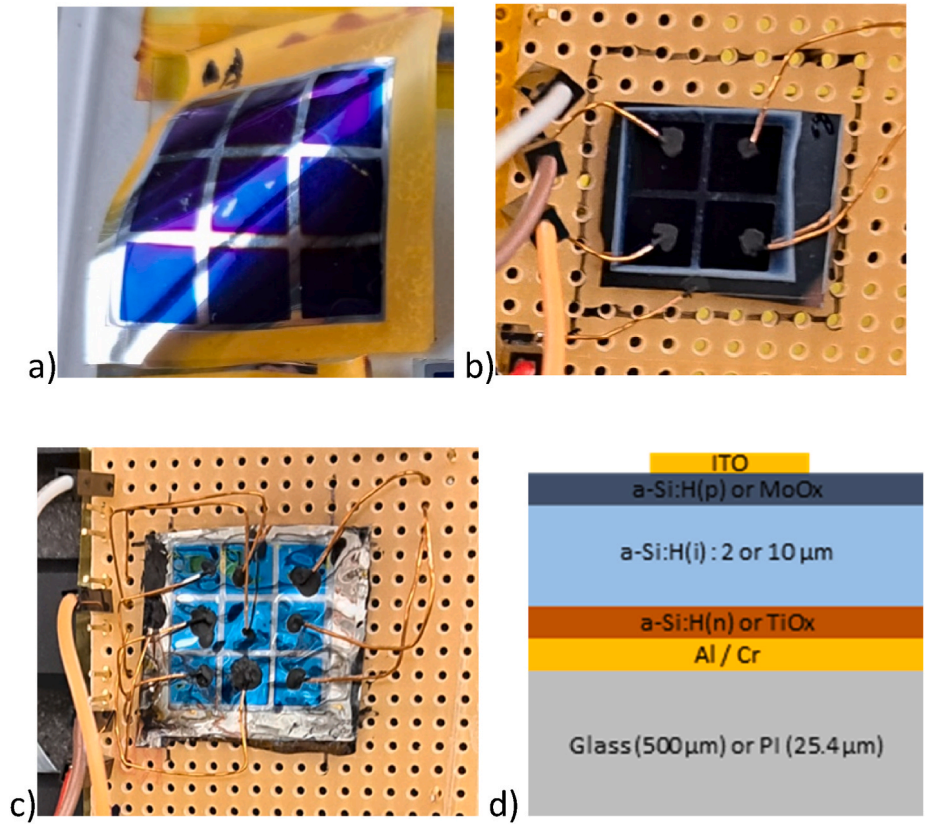

**Fig. 1.** (a) unconnected 2 μm n-i-p device on PI, (b) 10 μm n-i-p device on glass, (c) 2 μm device with selective contact on PI and (d) schematic cross-section of the devices.

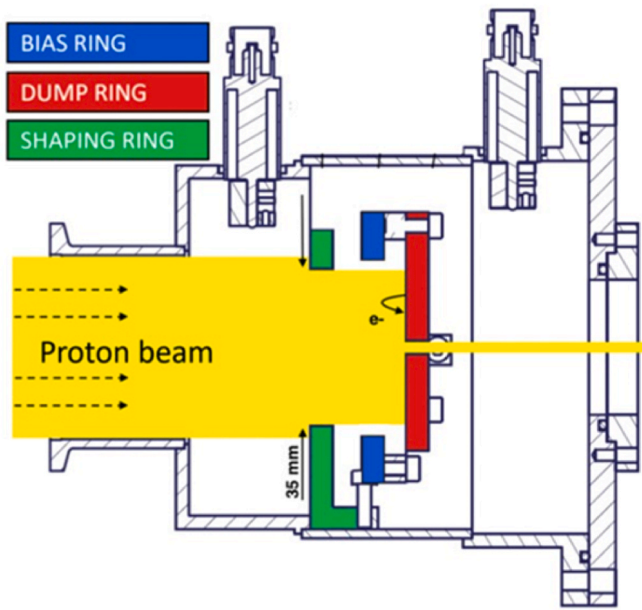

**Fig. 2.** Schematic view of the extraction end of the cyclotron research beam with the collimator (dump ring) use to define the beam size and measure the fluence.

The dose rate on the sample surface was then calculated from the proton flux, taking into account the proton beam energy reduced by the energy loss in the Al extraction window and the stopping power of Silicon. In the present configuration, the determined dose rate is a linear function of the collimator current. The overall uncertainty (≤5%) on the dose rate results from reading the proton beam current (1%) (Dellepiane et al., 2022) and from radiochromic film dosimetry. Note that these a-Si: H devices deposited on PI are almost tissue equivalent (the main contribution is given by the PI substrate, which is close to tissue equivalent).

### 2.3. Device characterisation

A first campaign of measurements at various dose rates with similar a-Si:H n-i-p detectors showed a slight degradation of the detector as a function of the total dose (or irradiation time). However, a good linearity as a function of dose rate was demonstrated for similar total doses for the same devices. To get a reliable evaluation of the linearity and radiation hardness the following procedure was adopted (pictured in Fig. 3).

To verify the linearity of the detectors, all samples were tested at several total doses by using short exposure times (of a few seconds), thus by varying the average dose rate. After each exposure, a waiting time of a few tens of seconds was applied to allow the sensor current to reach its initial (beam off) value (within a few percent). The latter was checked before any new beam exposure. Between these linearity tests, detectors were exposed to long exposure times (up to 30 min) with dose rates up to a few kGy/s. The sensitivity of the detectors at various total doses and as a function of time could then be evaluated. Note that, due to the indirect control of the dose rate of the proton beam, the range and selected values for the dose rates were different for all samples and measurement series.

As observed for several samples, the detector response tends to increase as a function of time when the beam is on at high dose rate. This observation will be discussed in section 3.3. To minimize this effect on the measurements, sensor current values were evaluated from a linear fit recorded in a short time window after the beam activation, between 0.5 s (to let the beam stabilize) and 1.5 s (to avoid a significant increase of the signal – see Fig. 3b). Note that the sensor current saturates after a long time, as observed in Fig. 3a.

## 3. Results

### 3.1. Device current as a function of dose rate

Fig. 4 shows the current of a 2 μm thick n-i-p a-Si:H device on glass as a function of the dose rate at three different reverse bias voltages. In these plots, the crosses indicate the current measured just after switching on the proton beam (see also section 3.3). Especially at high dose rates, the successive measurements (while the beam is kept on) exhibit a drift to higher current values or lower in the case of the measurements at 0 V. This effect was observed for different samples and will be explored in more detail in section 3.3. Considering only the first measurement points, a clear linear behaviour is observed at all bias voltages for this device. Moreover, the sensitivity (given by the slopes of the fitting line) is considerably reduced as the bias voltage is reduced, as a result of the decrease of the depletion region thickness. At 0 V, the latter is considerably reduced and should be more sensitive to the creation of defects in the intrinsic region of the detector. This may explain the reduced linearity observed compared to measurements at higher bias voltages.

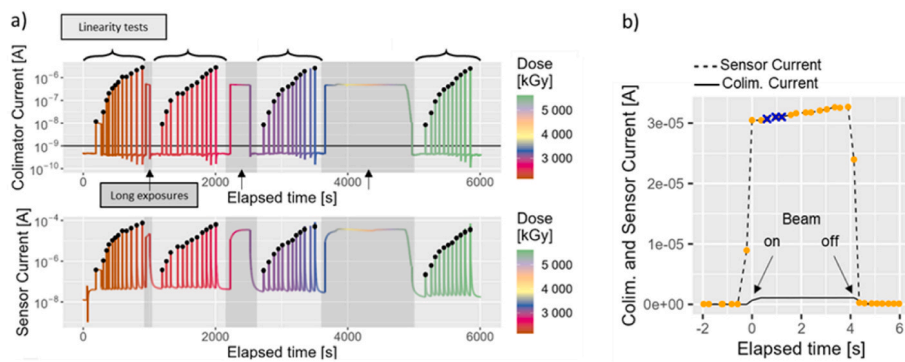

**Fig. 3.** a) Schematic exposure profile (collimator current as a function of the time) of the sample 2 μm n-i-p device on glass and the corresponding sample (sensor) current. The total dose received by the sample is indicated by the colour scale. Black dots represent measurement points for the evaluation of the detector sensitivity. b) Zoom on the elapsed time during which the beam is activated (as illustration). The crosses indicate the three data points selected for the linear fit of the sensitivity, 0.5–1.5 s after the beam was activated. Only three points were considered to avoid nonlinear increase observed at high dose rates.

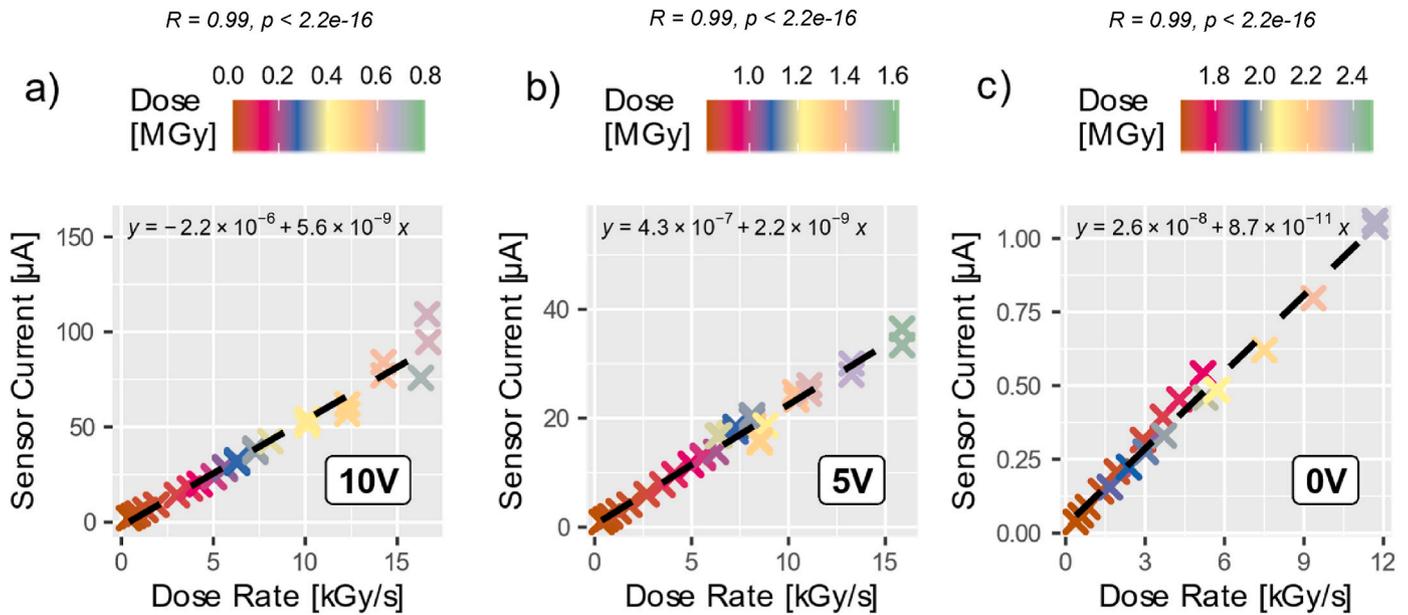

**Fig. 4.** Current as a function of dose rate of a 2 μm thick n-i-p a-Si:H device on glass at 3 different reverse bias voltages (a: 10 V, b: 5 V, c: unbiased – passive 0 V condition). The fit equation indicates the sensor sensitivity in nC/Gy as well as fit quality (R-squared and p-value). Uncertainties are estimated to be less than 5% on the dose rate and less than 2% on the sensor current.

Fig. 5 shows the current as a function of the dose rate of 2 μm thick a-Si:H devices with selective contacts on glass and on PI as a function of the dose rate at 10 V reverse bias voltages. As already observed for the n-i-p configuration, these devices also exhibit excellent linearity over the tested range of dose rates but with a slightly lower sensitivity (when deposited on the same type of substrate). As also observed in this figure, the sensitivity values of samples deposited on PI are lower than those deposited on glass. It is attributed to a lower device quality due to the roughness of the PI substrates resulting from the lamination process in the fabrication of the PI foils.

Fig. 6 reports the measurements on 10 μm thick devices. Only samples deposited on glass could be analysed due to the rather severe delamination observed on samples deposited on PI. The measured samples also exhibited large leakage currents that prevented the application of a sufficient high bias voltage to fully deplete the devices. At 10 V reverse bias voltage, the sensitivity is, therefore, even lower (2.1 nC/Gy, Fig. 6a) than on 2 μm thick n-i-p devices (5.6 nc/Gy). By increasing the bias voltage from 10 V to 20 V (Fig. 6 b and c) a doubling of the sensitivity is observed. However, this device with selective contacts remains still mostly undepleted as a bias voltage greater than 200 V is probably needed to achieve full depletion. It should be noted that the voltage needed for full depletion scales as the square of the thickness (Wyrsch and Ballif, 2016). All measurements are summarised in Table 1 (note that not all of them are plotted in figures).

As observed in the figures and Table 1, the linear fit quality is very satisfactory for all samples and bias voltage, except for the sample measured at 0 V, where the total dose during the experiments leads to deviations in the linearity. For all other samples, no deviation from linearity is observed within the measurement accuracy despite the increase in the total dose. One should note that the chosen methodology was not designed to minimize total dose increase during the experiments; exposure time for each single dose rate measurement could have been reduced considerably.

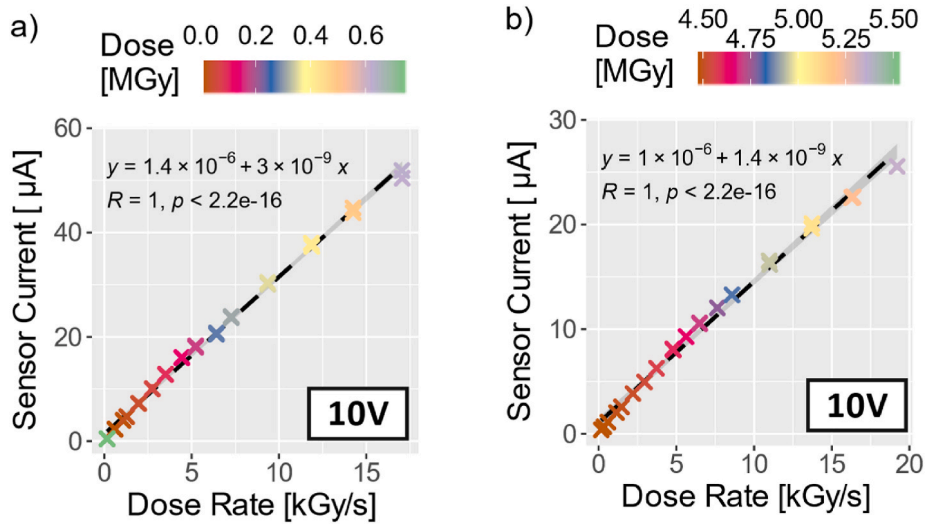

**Fig. 5.** Current as a function of dose rate of a 2 μm thick a-Si:H device with selective contacts on glass (a) and on PI (b) at 10 V reverse bias. Uncertainties are estimated to be less than 2% on the dose rate and less than 5% on the sensor current.

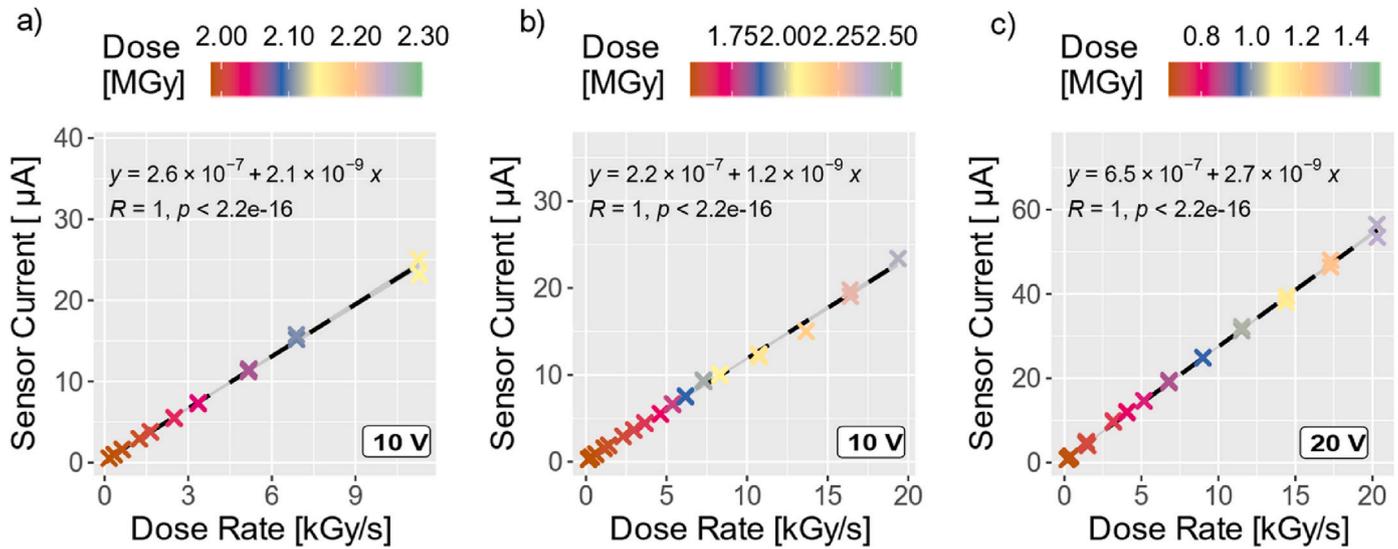

**Fig. 6.** Current as a function of dose rate of (a) a 10 μm thick n-i-p a-Si:H device on glass at 10 V and for (b) a 10 μm thick a-Si:H device with selective contacts on glass at 10 V and (c) 20 V reverse bias voltages. Uncertainties are estimated to be less than 5% on the dose rate and less than 2% on the sensor current.

**Table 1**
Sensitivity as a function proton dose rate of all samples at various reverse bias voltage values. The fit linearity (R-squared) for the sensitivity is also indicated as well as the total dose before and at the end of the last exposure used for the sensitivity determination. All sensitivity values were calculated for a dose rate range between 0 and 12–18 kGy/s.

| Sample contacts | Thickness [μm] | Substrate | Bias [V] | Sensitivity [nC/Gy] | R-squared | Initial/final total dose [MGy] |
|---|---|---|---|---|---|---|
| n-i-p | 2 | Glass | 10 | 5.6 | 0.99 | 0/0.8 |
|  |  |  | 5 | 2.2 | 0.99 | 0.8/1.6 |
|  |  |  | 0 | 0.087 | 0.99 | 1.6/2.5 |
| Selective | 2 | Glass | 10 | 3.0 | 1 | 0/1.0 |
|  |  |  | 5 | 1.6 | 1.0 | 1.3/2.0 |
|  |  |  | 0 | 0.17 | 1.0 | 2/2.8 |
| Selective | 2 | PI | 10 | 1.4 | 1.0 | 4.5/5.5 |
|  |  |  | 5 | 1.0 | 0.99 | 1.6-3.0 |
|  |  |  | 0 | 0.018 | 1.0 | 3.2/4.4 |
| n-i-p | 10 | Glass | 10 | 2.1 | 1 | 1.9/2.2 |
| Selective | 10 | Glass | 20 | 2.7 | 1.0 | 0.6/1.5 |
|  |  |  | 10 | 1.2 | 1.0 | 1.5/2.5 |

### 3.2. Device sensitivity as a function of total dose

Sensitivity values as a function of the total dose rate of a 2 μm thick n-i-p a-Si:H device on glass are plotted in Fig. 7. We see that the sensitivity decreases as a function of the total dose, but that a recovery is also observed when the sample is let in the dark at room temperature. This metastable effect will be discussed in more detail in section 4. The initial and subsequent sensitivities (40 days later) were determined in the same configuration and on the same device. The calibration of the proton beam dose rate at the detector's position with respect to the collimator current was also checked separately for the two measurement campaigns.

### 3.3. Device current drift

The observed current drift, briefly mentioned above (circle points in the plots of Figs. 4–6), is better seen in Fig. 8, which shows the magnitude of the drift scale as a function of the dose rate. We also observe that this phenomenon is more evident for a wider proton beam, even though the sample size is the same. The beam size is at least twice the size of the sample, which is fully irradiated. As the diode current is highly dependent on temperature, we believe that this drift originates in the heating of the sample. With a larger beam size, more energy is dissipated in the glass substrate and the drift is more evident. Such a hypothesis is also supported by the comparison between a sample deposited on glass and one deposited on PI, the latter exhibiting a reduced drift, as observed in Fig. 9. This figure also shows that devices with selective contacts seem to be less sensitive to such a thermal load. The observation that these devices exhibit, in some cases, a less pronounced rectifying behaviour (lousy diode) could be the reason.

Expanding the exposure time, especially at high doses, the sensor current will increase until an equilibrium is achieved when heating is compensated by heat dissipation. In these conditions, as observed in Fig. 3, a stable current is observed during these long exposures.

The 2 μm thick sample with selective contacts deposited on PI exhibited similar but unstable sensitivity as the ones deposited on glass in the initial state (low total dose) or after annealing. More investigation of the effect of PI substrates on the detector's performance is needed.

### 4. Discussion

All devices exhibit a remarkable linear current as a function of dose rate, independently of their depletion state. With the present methodology and irradiation conditions, this was observed as long as the total dose did not change significantly. In the present work, within linearity tests, the total dose change was kept below 1 MGy and, for most of the measurement points, well below 0.5 MGy.

The material a-Si:H is known to be metastable (Wyrsch and Ballif, 2016; Street, 1991). Defects are created by recombination events, which may take place as soon as electron-hole pairs are created independently of the pair creation origin (light or particles) (Street, 1991; Frietzsche, 2001). These defects can then be thermally annealed, and this process takes place even at room temperature, but over much longer times. Additionally, defects can also be created directly by energetic particles, but the amorphous nature of the material and the large hydrogen content (ca. 10 atomic percent) efficiently passivate those defects. The metastable behaviour observed in this study (cf. Fig. 8) is very similar to the ones already observed earlier in irradiation experiments of devices with intense and energetic proton beams (Wyrsch et al., 2006).

In the context of dosimetry for radiation therapy, including FLASH

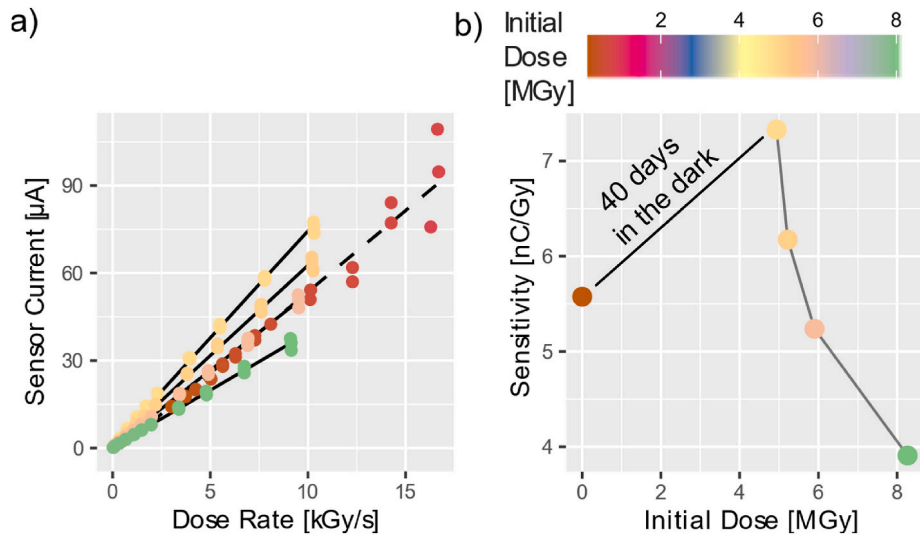

**Fig. 7.** (a) Current as a function of dose rate of a 2 μm thick n-i-p a-Si:H device on glass at 10 V reverse bias voltage as a function of the dose rate. The sensitivity is also plotted as a function of the total initial dose (b). It should be noted the increase in sensitivity from the first measurement series (Fig. 4) to a subsequent irradiation campaign. In between the device was stored 40 days in the dark.

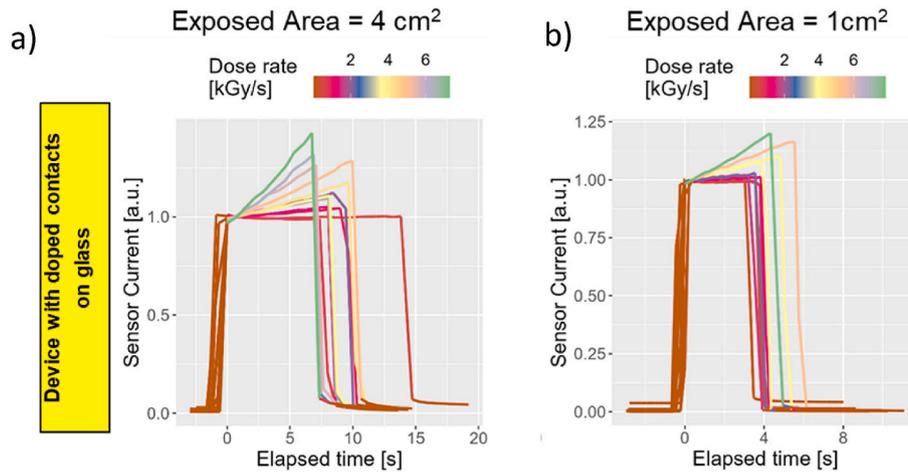

**Fig. 8.** Normalized sensor current as a function of time following the switching on of the proton beam as measured for a 2 μm thick n-i-p a-Si:H device on glass (10 V reverse bias) for two openings of the collimator resulting in a proton beam of (a) 2 × 2 cm$^2$, and (b) 1 × 1 cm$^2$, respectively.

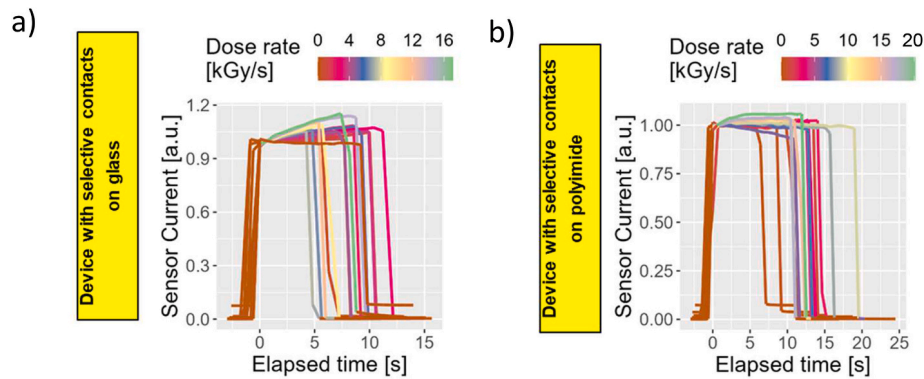

**Fig. 9.** Normalized sensor current as a function of time following the switching on of the proton beam as measured for a 2 μm thick a-Si:H device with selective contact deposited on glass (a) and on PI (b) for 10 V reverse bias with a proton beam of 1x1 cm$^2$.

therapy, this metastability and changes of sensitivity do not seem to be a critical aspect. Typically, for this novel radiotherapy modality, doses from a few Gy up to tens of Gy are delivered in sub-second times (Esplen et al., 2020)(Romano et al., 2022). Therefore, a detector should not receive more than a few kGy per day. As a self-annealing of the device also takes place, we expect to have a stable operation over a very long time (months to years).

The drift of the current observed as a function of time at high-dose rates is not surprising, as a dose rate of 10 kGy/s corresponds to the absorption of a power of 10 W/g. We can then expect, as observed, that a detector deposited on a 500 μm thick glass will absorb a much larger power than when deposited on 25.4 μm PI. As also observed, devices with charge-selective contacts are less sensitive to temperature change. We attribute this effect, for the present devices, to a less rectifying behaviour of these devices, which in some cases exhibit a rather leaky capacitor behaviour than a diode one. The use of such devices deposited on thin PI substrates can be advised for dosimetric measurements of long exposures.

As already observed and underlined, partial depletion does not affect the linear behaviour of the device both for n-i-p configuration as well as for charge selective contact. Even in unbiased devices (passive situation) an almost perfect linearity is observed, while the sensitivity is considerably reduced (cf. Table 1). This reduction is expected, as transport and charge collection in a-Si:H devices is controlled by drift and diffusion processes can be neglected. Therefore, only the device volume where an electric field is present (i.e. in the depletion region) will contribute to charge collection. In n-i-p devices, regions close to the n-i and p-i interface will exhibit band bending and, as a consequence, an internal field, contributing thereby to the charge collection. Band bending close to the interface with charge-selective contacts will play a similar role. In thick devices, only a limited part of the volume contributes and thus limits the potential sensitivity, unless the device can sustain high-bias voltage values necessary to achieve full depletion. Using an unbiased device simplifies the measurement setup at the cost of a much-reduced sensitivity and a larger effect of total dose on the latter to keep a linear behaviour between device output and dose rate.

## 5. Conclusions

a-Si:H devices in n-i-p configuration or with selective contacts demonstrate remarkable linear behaviour as a function of dose rates in the range 0–20 kGy/s. Devices in passive configuration (unbiased) exhibit similar linear behaviour despite a considerably reduced sensitivity. For this type of application, very thick devices do not offer improved sensitivity as they are more difficult to fully deplete.

The metastability of a-Si:H is well known and is also observed here for proton irradiations. The experimental conditions of external beams radiation therapy result in a much lower total dose delivered per day to the detectors than that used in this study. For this reason, a-Si:H is a promising material for real-time dosimetry in radiation therapy, including the novel FLASH therapy. Such devices could be used both for quality assurance, but also for real-time beam monitoring. As already reported in the field of particle detection (Wyrsch and Ballif, 2016) detectors with active areas as small as a few square micrometers can be fabricated.

However, more effort is still needed to fully understand their potential and limits. While these detectors are also expected to perform linearly at much lower doses (as anticipated from light detection experiments), further testing on a broader range of proton irradiation should also be performed. This would allow for the use of such devices for both conventional and FLASH radiotherapy. Application of these detectors should, in principle, be possible for other types of particles (ions or electrons) (Wyrsch and Ballif, 2016) but would require specific validation.

For practical applications in proton radiotherapy, the total dose is not expected to affect much the linearity of the detectors. However, for more demanding applications, regular recalibration of the detector may be needed as the sensitivity may change as a function of the detector's history. Alternatively, an annealing process to recover the material defect equilibrium could be implemented. Note that the time needed to recover the initial performance will depend on the device itself, history, and annealing condition (mainly the device temperature).

Devices deposited on Kapton foils, while offering the benefit of flexibility to perform dosimetric measurements on curved shapes or on the human body in particular, exhibit less interaction with the proton beam, leading to minimal heating of the device and, as a consequence, much lower signal drift. On this aspect, devices with charge-selective contacts have been seen to be less dependent on temperature change. However, further studies are needed to understand the effect of the substrate on the detector's performance.

Another important aspect of detectors for FLASH therapy is their device response time. a-Si:H photodiodes (similar to the detectors characterized here) have response time values below 1 μs (Shen et al., 1995). While this response should be tested in the context of dosimetry, we do not expect significant constrain from a-Si:H.

Further studies are needed to define the optimal device thickness of proton beam dosimetry to better understand the operation of devices with charge-selective contacts and to identify the pros and cons of this technology.

## CRediT authorship contribution statement

**Nicolas Wyrsch:** Writing – original draft, Validation, Supervision, Project administration, Methodology, Investigation, Funding acquisition, Formal analysis, Data curation, Conceptualization. **Luca Antognini:** Writing – review & editing, Visualization, Validation, Methodology, Investigation, Formal analysis, Data curation, Conceptualization. **Christophe Ballif:** Supervision. **Saverio Braccini:** Writing – review & editing, Supervision. **Pierluigi Casolaro:** Writing – review & editing, Methodology, Investigation, Conceptualization. **Sylvain Dunand:** Investigation, Conceptualization. **Alexander Gottstein:** Investigation. **Matt Large:** Methodology, Investigation. **Isidre Mateu:** Investigation, Conceptualization. **Jonathan Thomet:** Writing – review & editing, Formal analysis.

## Declaration of competing interest

The authors declare the following financial interests/personal relationships which may be considered as potential competing interests:

Nicolas Wyrsch reports financial support and administrative support were provided by Ecole Polytechnique Fédérale de Lausanne. If there are other authors, they declare that they have no known competing financial interests or personal relationships that could have appeared to influence the work reported in this paper.

## Data availability

Data will be made available on request.

## Acknowledgments

The authors acknowledge the financial support of the Swiss National Science Foundation under the grants 200021_212208/1 and CRSII5_180352, and of the Bern Center for Precision Medicine (BCPM) of the University of Bern.

## References

Anders, J., et al., 2022. A facility for radiation hardness studies based on a medical cyclotron. J. Instrum. 17, P04021 https://doi.org/10.1088/1748-0221/17/04/P04021.

Braccini, S., et al., 2023. A two-dimensional non-destructive beam monitoring detector for ion beams. Appl. Sci. 13, 3657. https://doi.org/10.3390/app13063657.

Bracini, S., 2013. The new Bern PET cyclotron, its research beam line, and the development of an innovative beam monitor detector. AIP Conf. Proc. 1525, 144–150. https://doi.org/10.1063/1.4802308.

Casolaro, P., Campajola, L., Di Capua, F., 2019. The physics of radiochromic process: one calibration equation for all film types. J. Instrum. 14, P08006 https://doi.org/10.1088/1748-0221/14/08/P08006.

Casolaro, P., et al., 2022. Time-resolved proton beam dosimetry for ultra-high dose-rate cancer therapy (FLASH). In: Proc. Of 11th Int. Beam Instrum. Conf., 519. https://doi.org/10.18429/JACoW-IBIC2022-WE3C2.

Dellepiane, G., et al., 2022. Cross section measurement of terbium radioisotopes for an optimized 155Tb production with an 18 MeV medical PET cyclotron. Appl. Radiat. Isot. 184, 110175 https://doi.org/10.1016/j.apradiso.2022.110175.

Diffenderfer, E.S., et al., 2022. The current status of preclinical proton FLASH radiation and future directions. Med. Phys. 49, 2039. https://doi.org/10.1002/mp.15276.

Esplen, N., Mendonca, M.S., Bazalova-Carter, M., 2020. Physics and biology of ultrahigh dose-rate (FLASH) radiotherapy: a topical review. Phys. Med. Biol. 65, 23TR03 https://doi.org/10.1088/1361-6560/abaa28.

Frietzsche, H., 2001. Development in understanding and controlling the staebler-wronski effect in a-Si:H. Annu. Rev. Mater. Res. 31, 47. https://doi.org/10.1146/annurev.matsci.31.1.47.

Grimani, C., et al., 2023. A hydrogenated amorphous silicon detector for Space Weather applications. Astrophys. Space Sci. 368, 88. https://doi.org/10.1007/s10509-023-04235-w.

Hughes, J.R., Parsons, J.L., 2020. FLASH radiotherapy: current knowledge and future insights using proton-beam therapy. Int. J. Mol. Sci. 21, 6492. https://doi.org/10.3390/ijms21186492.

Jolly, S., Owen, H., Schippers, M., Welsch, C., 2020. Technical challenges for FLASH proton therapy. Phys. Med. : PM : an international journal devoted to the applications of physics to medicine and biology : official journal of the Italian Association of Biomedical Physics (AIFB) 78, 71–82. https://doi.org/10.1016/j.ejmp.2020.08.005.

Large, J.L., et al., 2023. Hydrogenated amorphous silicon high flux x-ray detectors for synchrotron microbeam radiation therapy. Phys. Med. Biol. 68, 135010 https://doi.org/10.1088/1361-6560/acdb43.

Mascia, A.E., Daugherty, E.C., Zhang, Y., et al., 2023. Proton FLASH radiotherapy for the treatment of symptomatic bone metastases: the FAST-01 nonrandomized trial. JAMA Oncol. 9, 62–69. https://doi.org/10.1001/jamaoncol.2022.5843.

Matuszak, N., Suchorska, W.M., Milecki, P., Kruszyna-Mochalska, M., Misiarz, A., Pracz, J., Malicki, J., 2022. FLASH radiotherapy: an emerging approach in radiation therapy. Rep. Practical Oncol. Radiother. : journal of Greatpoland Cancer Center in Poznan and Polish Society of Radiation Oncology 27, 344–351. https://doi.org/10.5603/RPOR.a2022.0038.

Menichelli, M., et al., 2021. Fabrication of a hydrogenated amorphous silicon detector in 3-D geometry and preliminary test on planar prototypes. Instruments 5 32. https://doi.org/10.3390/instruments5040032.

Menichelli, M., et al., 2023. X-ray qualification of hydrogenated amorphous silicon sensors on flexible substrate. 9th International Workshop on Advances in Sensors and Interfaces (IWASI) 190. https://doi.org/10.1109/IWASI58316.2023.10164611.

Potkins, D.E., et al., 2017. A low-cost beam profiler based on cerium-doped silica fibers. Phys. Procedia 90, 215. https://doi.org/10.1016/j.phpro.2017.09.061.

Romano, F., Bailat, C., Jorge, P.G., Lerch, M.L.F., Darafsheh, A., 2022. Ultra-high dose rate dosimetry: challenges and opportunities for FLASH radiation therapy. Med. Phys. 49 (7), 4912–4932. https://doi.org/10.1002/mp.15649.

Shen, D., Kowel, S.T., Eldering, C.A., 1995. Amorphous silicon thin-film photodetectors for optical interconnection. Opt. Eng. 34, 881. https://doi.org/10.1117/12.190450.

Street, R., 1991. Hydrogen diffusion and electronic metastability in amorphous silicon. Physica B 170, 69. https://doi.org/10.1016/0921-4526(91)90108-Q.

Vozenin, M.C., Hendry, J.H., Limoli, C.L., 2019. Biological benefits of ultra-high dose rate FLASH radiotherapy: sleeping beauty awoken. Clin. Oncol. 31, 407–415. https://doi.org/10.1016/j.clon.2019.04.001.

Wilson, J.D., Hammond, E.M., Higgins, G.S., Petersson, K., 2020. Ultra-high dose rate (FLASH) radiotherapy: silver bullet or fool's gold? Front. Oncol. 9, 1563. https://doi.org/10.3389/fonc.2019.01563.

Wyrsch, N., Ballif, C., 2016. Review of amorphous silicon based particle detectors: the quest for single particle detection. Semicond. Sci. Technol. 31, 103005 https://doi.org/10.1088/0268-1242/31/10/103005.

Wyrsch, N., et al., 2004. Thin-film silicon detectors for particle detection. Phys. Status Solidi 5, 1284. https://doi.org/10.1002/pssc.200304329.

Wyrsch, N., et al., 2006. Radiation hardness of amorphous silicon particle sensors. J. of Non-Cryst. Sol. 352, 1797. https://doi.org/10.1016/j.jnoncrysol.2005.10.035.